# Software Architecture for Operation and Use of Quantum Communications Networks

Dinesh C. Verma
IBM TJ Watson Research
Yorktown Heights, NY,
U.S.A.
dverma@us.ibm.com

Eden Figueroa
Dept. of Physics
State University of NY
Stonybrook, NY, U.S.A
eden.figueroa@stonybrook.edu

Gabriella Carini
Instrumentation Division
Brookhaven National Lab
Upton, NY, U.S.A.
carini@bnl.gov

Mark Ritter
IBM TJ Watson Research
Yorktown Heights, NY,
U.S.A.
mritter@us.ibm.com

*Abstract—* Quantum Communications Networks using the properties of qubits, namely state superposition, no-cloning and entanglement, can enable the exchange of information in a very secure manner across optical links or free space. New innovations enable the use of optical repeaters as well as multi-cast communication in the networks. Some types of quantum communications mechanisms can be implemented at room-temperature instead of requiring super-cooled systems. This makes it likely that business impact from quantum communications will be realized sooner than that from quantum computers.

Quantum networks need to be integrated into the ecosystem of currently deployed classical networks and augment them with new capabilities. Classical computers and networks need to be able to use the new secure communication capabilities offered by quantum networks. To provide this interoperability, appropriate software abstractions on the usage of quantum networks need to be developed. In this paper, we examine what the type of software abstractions quantum networks can provide, and the type of applications that the new abstractions can support.

*Keywords—* *Quantum networks, communications software, network software, communication abstractions*

## I. INTRODUCTION

Quantum Communications Networks [1], [2] enable secure information exchange and have been demonstrated over optical links as well as free space communication. Quantum communications can be supported on a complex network topology using optical repeaters [3], [4] and they can support multi-cast communication [5], [6]. Quantum communications can be implemented at room-temperature [7], [8] instead of requiring super-cooled systems, which makes realizing the business impact from quantum communications at a time-scale which may be faster than the impact of quantum computers themselves.

To deal with real-world issues such as information loss and transmission errors, quantum networks are usually accompanied by a classical network, which is used to support and help in the operations of the quantum network. This provides for a new type of communication network which combines the best attributes from both quantum and classical networks to improve the nature of what is possible between one or more pair of communicating computers.

Quantum networks provide a greenfield space, giving us an opportunity to rethink how networks ought to be controlled, operated and used. The last few decades have seen many different types of networks [9], [10], including but not limited to circuit-switched telephony networks [11], packet-switched Internet [12], Broadband Integrated Services Digital Networks (B-ISDN) or Asynchronous Transmission Mode (ATM) networks [13], Multiprotocol Label Switching (MPLS) networks [14], device-oriented third generation (3G) and fourth generation (4G) cellular networks [15], software oriented fifth generation (5G) cellular networks [16], [17], centrally controlled Software Defined Networks [18], Supervisory Control And Data Acquisition (SCADA) networks [19] etc. The success and failures of these networks have also taught the technical community important technical lessons and best practices in the control, operation and management of computer networks. It is our goal to combine the best lessons learnt from the design of control, management and operations of the past type of networks to design a good control software architecture for quantum networks.

In this paper, we propose such a software architecture that covers the topics of using, controlling and managing quantum communications networks. This architecture will cover the data, control and management planes required for quantum communication networks.

We believe this is the first paper to explore how software ought to be developed for Quantum-enabled Internet. While there are papers that have explored subjects such as using software defined networking to control quantum nodes [20], [21], developing simulators for quantum networks [22]–[24], and modeling of quantum protocol performance [25], [26], we have not come across papers that have explored subjects involved with the development of software on classical computers to exploit quantum networks. To enable this goal, we propose new communication abstractions that ought to be supported by a software development kit (SDK) using the capabilities of a quantum network and move towards practical application software on a quantum-enabled Internet. As a first attempt towards this topic, we do not presume to assert that the abstractions we introduced are the final one, but strive to initiate a discussion in the technical community on the subject of the right software abstraction to use and exploit quantum communication networks.

After making a few general observations on the lessons learnt from software architectures from existing types of classical networks, we discuss the various configurations that a quantum network can be used for. We argue that we ought to focus initial discussion of software development on classical computers communicating over quantum networks. We subsequently discuss the abstractions that will be useful to develop new software applications using quantum communications. This abstraction provides the data plane architecture for quantum networks, and we briefly cover the software layers needed for the control and management plane of quantum networks. We conclude by discussing some software applications that can benefit from the abstractions we propose.

## II. GENERAL OBSERVATIONS

Years of designing computer networks of many flavors have led to some lessons and best practices in the design of software for computer networks. As we attempt the design for the software stack to operate a new class of networks, it is instructive to recap some of the best practices learnt from existing networks.

The first of the extremely successful lessons is the importance of *layering*. The concept of designing computer networks as consisting of many layers, with each layer solving one problem, relying on layers above or below it to address other problems, and exporting a canonical interface with a well-defined set of abstractions has proven its value in the implementation of all types of networks. It has also allowed different types of networks to be layered in new and unique ways. ATM networks [13], which were designed to be complex global standardized networks became a data-link layer to carry IP packets of the Internet due to economic and business reasons. SCADA networks [19], designed initially with a hardware set of components in mind started to get layered on top of IP networks. User-level implementations of network protocols such as IP were done to bypass the difficulties of implementing software in operating system kernels instead of in the user space. It follows implicitly that we should design the software of the quantum network as a layer which exports a well-defined set of abstractions to its user.

The second lesson from the development of various protocols has been in the separation of the functions in *three different planes of data, control and management*. Circuit switched networks and current cellular networks have well-defined protocols delineated in these tree different planes. The Internet did not have a control plane for a long period but the emergence of protocols like SIP [27] and the advantages of managing Ethernet switches using Software Defined Networking [18] reinforced the value of having a control plane within the network. On a broad basis, the following is the separation between these three planes of network:

- **Control Plane**: The control plane suite of functions is responsible for setting up the initial mechanism for establishing communication. In a circuit switched network, this established the end-to-end circuit. In a packet switched network, this capability is not needed. Nevertheless, having a control plane to orchestrate any steps required to ensure smooth communication would be useful for any type of network.

- **Management Plane**: The management plane suite of functions is responsible for managing any errors that happen during the operation of the network. This requires observing events in the network, detecting problems that may be present, and fixing those problems. The problems can either be reported to a human administrator for appropriate action, or automated actions for fixing the problems can be taken, e.g. using an automation engine or an Artificial Intelligence based engine.

- **Data Plane**: The data plane carries the information flowing on the network, and is responsible for the aspects such as reliable transmission of information, security of the information, and supporting the communication abstraction offered by the network.

Quantum communications network have a strong flavor of being a circuit-switched network, and hence it would make sense to define a software architecture that includes all three planes for their operation.

The design of each of these three planes could consist of one or more layers, i.e. layering may help us design a better control and management plane for the operation of quantum networks.

## III. QUANTUM NETWORK CONFIGURATIONS

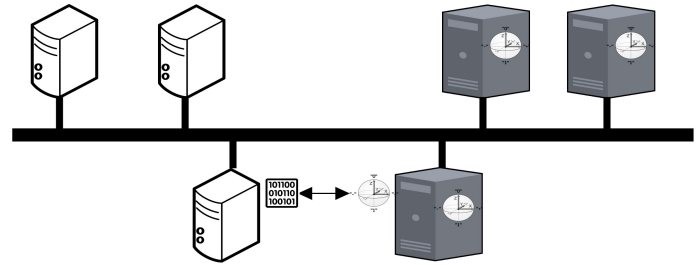

*Figure 1. Configurations of Quantum Networks*

Before describing the design of the software in each of the three planes of control, management and data, it would be instructive to look at the usage configurations of a quantum network. A quantum network can be used in one of the following three configurations:

1) Quantum-enabled Configuration: The quantum network is used to interconnect one or more classical computers together. To the classical computers, the quantum network is another channel to exchange information among themselves.

2) Full Quantum Configuration: The quantum network is used to interconnect one or more quantum computers together. To the quantum computers, the quantum network is a channel to exchange qubits among themselves. Quantum computers may choose to exchange the entanglement state of the qubits, as opposed to an actual exchange of qubits.

*3) Mixed Configuration:* The quantum network is used to interconnect computers, some of which may be classical computers while others may be quantum computers. To the classical computers, the quantum network provides a way to get the qubits from the quantum computers on the network, which they can read and collapse the information to a binary number. The quantum computers, the quantum network provides a way to transform a classical number into a set of qubits.

The mixed configuration would match the vision of Quantum Internet [28], but the quantum-enabled configuration and full quantum configuration are likely intermediary steps towards the attainment of the vision. A mixed configuration can be supported by means of a quantum and a classical computer, both being present in the network and the conversion between qubits and classical bits be done outside the quantum communication network. In effect, the mixed configuration quantum network is separable into two segments, one where a group of classical computers are using it, and the other where a group of quantum computers are using it. The exchange between them two segments can happen as shown in Fig. 1. As a result, for the purpose of developing software abstractions, we can safely assume that for all practical purposes, there are only two configurations of the network - the quantum-enabled configuration and the quantum configuration.

When we consider a full quantum configuration, there is a physical constraint that needs to be taken into account. Quantum computers, in most of the common designs, operate at temperatures close to absolute zero. There are some explorations to create room-temperature quantum computing [29], but they are at a relatively early stage of development. On the other hand, quantum networks operating over wide area need to operate at room temperatures. Since the energy, entropy and entangled state of qubits can experience a significant change when moving from a super-cooled environment to a room-temperature environment, interconnecting a supercooled quantum computing system with a room-temperature quantum network remains a difficult problem at the current time. While we as a research community look forward to a solution to that problem in the future, and we can consider the theory of such distributed quantum computing environment, from the perspective of software development, it seems more prudent to focus on configurations that are viable currently.

Therefore, we will assume in the rest of the paper that we are operating in the quantum-enabled configuration.

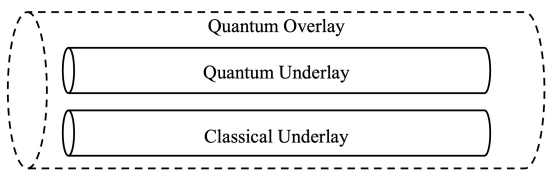

Fig. 2. The Abstracted Representation of a Quantum Network

We now discuss the abstractions that the quantum-enabled configuration ought to support. Given the current state of quantum networks, we can model the overall operation of the quantum network as an abstraction shown in Fig. 2. The bigger dashed pipe shows the representation of the quantum network, which we will refer to as the quantum overlay in order to avoid confusion with the network pipe that actually performs the transfer of qubits, which is the quantum underlay. The quantum underlay is supported by a classical underlay network. In order to communicate, both the services of the quantum underlay and the classical overlay need to be used. The combined pair of quantum underlay and the classical underlay provide the abstracted network representation, which is the quantum overlay.

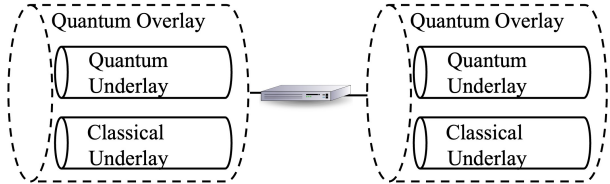

Fig. 3. Overlay repeated at each node

The quantum overlay provides the perspective of one link, and this link can be converted into the concept of a path or a network in one of two manners. The first approach in which this can be converted is shown in Fig. 3. In this case, the abstract quantum overlay is defined on a link by link basis. The control and management plane are defined to compose multiple quantum overlays together. The end to end path consist of a concatenation of multiple quantum overlays. The computer shown in the middle of Fig. 3 is a trusted relay which is a classical computer and reads/writes information across the two quantum overlays that it connects.

An alternate architecture is that of creating a more complex underlay. This approach assumes that the underlays create their own complete networks. This definitely holds true for the underlay of a classical computer network. In these cases, the network can be configured by means of a variety of control and management protocols. The same scheme for quantum networks, however, is not as well developed. This approach is shown in Fig. 4. The quantum underlay is connected by quantum repeaters, while the classical underlay is connected by traditional routers or switches.

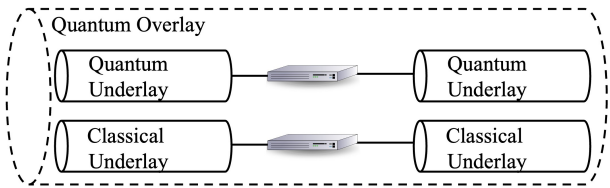

Fig. 4. Overlay Abstraction using Complex Underlays

A hybrid approach, in which the classical network is assumed to be present and interconnecting all sites which are used with the quantum network, with the quantum overlay being used at each location can also be used. This approach provides a mixed abstraction which is the one we prefer to follow. This allows the quantum overlay to be configured in a manner which need not configure the classical network, and only configure and manage the quantum network. This approach is shown in Fig. 5.

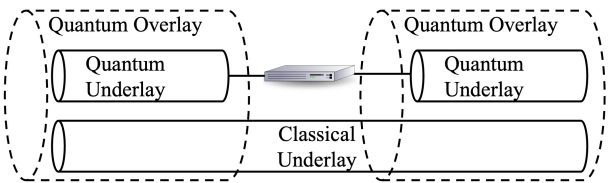

Fig. 5. Overlay Abstraction using Complex Classical Underlay

While the hybrid representation does not appear very clean from a conceptual representation, it does provide a pragmatic approach where we can focus on the control and management of quantum aspects of the network, instead of duplicating an already existing rich set of activities in the classical network space. With a hybrid representation, the control and management plane for quantum overlays can focus completely on the operation of the overlay and quantum underlay, assuming that the classical network is available and connect all the communicating nodes. For the discussion of the control and management architecture of the quantum overlays in the next section, we assume that the hybrid configuration is being used.

## IV. ABSTRACTIONS

Borrowing from the different personas that are used in classical communication networks, we can identify three types of personas that are involved with quantum communication networks. These personas include:

- **The User**: The user persona is the entity using the network for communication. In most cases, the user persona would be the application software that calls upon the quantum network to perform its task. If we consider the quantum network as one aggregated layer in the networking stack, the user is the software that is running on the layer above the quantum network layer.

- **The Provider**: The provider persona is the layer that sits below the quantum network

- **The Configurator**: The configurator persona is the entity that is responsible for getting an instance of the quantum network installed and ready for operation. The task of the configurator is to provide the control mechanisms that will ensure that the quantum network instance is up and operational.

- **The Administrator**: The administrator persona is the entity that is responsible for ensuring that the network is operational, that any errors happening in the network are handled correctly.

These three personas are effectively the users of the data plane, the control plane and the management plane for the quantum overlay that we have discussed in Section II. Let us now look at the abstractions that we can offer to each of these personas from the quantum overlay perspective.

### A. Data Plane Abstraction

The quantum overlay in the model we are describing leverages the two underlays - namely the quantum underlay and the classical underlay. As a default, the abstraction offered by the quantum overlay can be a superset of any of the abstractions offered either of the two overlays.

The traditional networking layer of the 7-layer OSI architecture [30] offers the abstractions of a circuit or a datagram. A circuit provides the abstraction of an end-to-end connection between the sender and the receiver. Information sent at one end of the circuit is delivered to the other end of the circuit. The circuit could be physical or virtual – the latter being more common given the ubiquity of the Internet Protocol. The circuit may have some error rate, but maintains the sequence of packets. A datagram provides the abstraction of a packet which is transferred between the end-points provided a destination of the packet. Delivery may be out of sequence, and may be lossy.

While we can export the same abstractions at the quantum overlay level, it does not leverage the capabilities of the quantum underlay. The quantum network due to the qubit properties of no-cloning, is more suitable for a circuit abstraction instead of a datagram abstraction. We propose the following abstractions to be offered by the quantum overlay which combine the strengths of the quantum underlay and the classical underlay:

- **Secure Lossy Datagram Circuit**: The secure unreliable circuit takes a datagram from the sender and delivers it securely to the receiver. Security is obtained by establishing a secure key using a Quantum Key Distribution (QKD) protocol [31], which is then used to encrypt the datagram. The secure circuit would have an key refresh interval parameter which would result in the renegotiation of the secure key after a given number of datagrams are transmitted, or after the lapse of a time-period. The secure circuit does not guarantee reliability of the circuit.

- **Secure Reliable Datagram Circuit**: The secure reliable circuit offers security as well as the reliability of datagrams that are delivered. The abstraction offered is still that of a datagram bundle, but information is guaranteed to be delivered. The delivery need not preserve sequence.

- **Secure Reliable ByteStream**: The secure reliable ByteStream offers security, reliability and in-sequence delivery. This allows the creation of a bytestream abstraction.

- **Synchronized Random number generator**: the secure random number generator provides an abstraction where a call of either side of the quantum overlay results in the generation of the same random number at both sides. The synchronized random number generated could be the key generated by QKD protocols which leverage both the classical and the quantum underlay. We would emphasize that the term synchronized here refers to the fact that the same number is seen by all parties, and does not imply any type of time-synchronization.

The combination of a datagram abstraction with a circuit abstraction may appear contradictory at first glance, but the usage model with quantum networks enables a logical way to understand this combination. The quantum underlay provides the circuit analog, while the ability to chunk information into logical units provides the concept of a datagram.

The first three abstractions provide the analogue of the abstractions offered by transport layer security (TLS) [32] over Universal Datagram Protocol(UDP) [33], TLS with Real-time Transport Protocol(RTP) [34], and TLS with Transmission Control Protocol (TCP) [35] in classical computer networks. The quantum overlay thus provides abstractions which are at transport and session level of the OSI-architecture [30].

The fourth abstraction exploits QKD protocols such as BB84 [36] or any other QKD protocol [31] to create a random number generator. This random generator can be used for other programs in a classical computer, such as the use in applications such as simulated annealing, genetic algorithms, or to provide seeds for a variety of Monte Carlo simulations. Having a synchronized random number at different end-points can enable the exploration of different spaces in Monte Carlo simulations more effectively. Unlike the first three abstractions, which rely on having a circuit traditionally associated with two end-points, the synchronized random number generator can be easily extended to the case of multicast quantum networks.

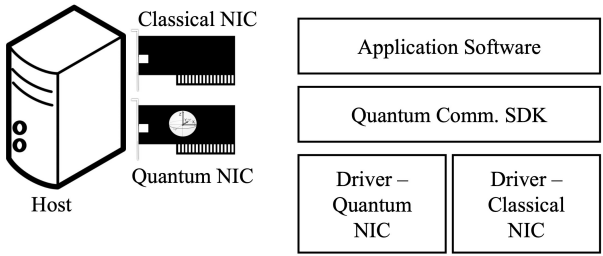

Fig. 6. Hardware and Software for supporting the Abstractions

Each of these abstractions can be implemented and provided by a software development kit (SDK) running on a classical computer. The assumptions of the hardware and software required for supporting this SDK are shown in Fig. 6. The host that runs the software have two network interface cards (NICs), one supporting a classical network and the other interfacing with a quantum network. The design of the quantum network NIC can be done in a variety of ways, just like many different kinds of classical NICs are available in the market today. The Quantum Comm. SDK exports the four abstractions described above to the application software that invokes it. It implements its abstractions by invoking the commands supported by the software drivers of the two NICs, using the capabilities of both the quantum network and the classical network.

The SDK allows computer software to be developed for quantum networks without waiting for a complete development of quantum computers. This SDK can be implemented and offered as a user-level library, just like the support for TLS protocol [37] is implemented in current classical computers.

### B. Control Plane Abstraction

The control plane is responsible for configuring the operation of the quantum overlay. The control plane for the quantum overlay needs to configure both the quantum underlay and the classical underlay in order for the quantum overlay to support the data plane abstractions. The control plane capabilities for a classical network are well developed. The additional work is to augment the control plane capabilities for the quantum network as well.

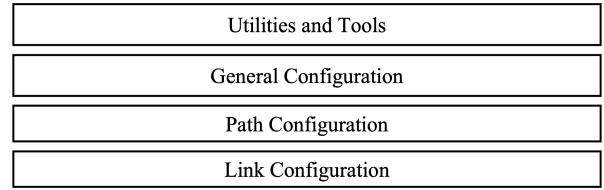

Fig. 7. The Control Plane for a Quantum Network

In order to perform its task, the control plane needs to implement the abstractions of a link, a path and a network. The link provides for a single instance of the quantum overlay (or the quantum underlay) since we are opting to go for the more hybrid approach for implementing the quantum overlay. The path is a collection of links, while the network consists of an conglomerate of paths and networks.

In order to control the network, a set of layered abstractions as shown in Fig. 7 ought to be implemented in the control layer. The link and path configurations are used to establish the specific communication needs to set up the quantum overlay abstraction. These would configure the different attributes so that the quantum underlay network is able to exchange information, and appropriate configuration information is also provided and exchanged for a quantum network.

The topmost layer of the control plane architecture is the set of utilities and tools that will be needed. These utilities and tools include the software that can perform tasks such as autoconfiguration, and the generation of policies so that policy based configuration [38], [39] can be supported. Policy based management has been used to configure many different types of networks, and would be a natural approach to configure and control quantum overlays/underlays.

A policy based approach can be used with both decentralized control mechanisms, e.g. how distributed routing protocols control the operation of the network, as well as with centralized control mechanisms such as SDN [18]. Some papers have explored using the centralized control scheme for quantum networks [20]. Both centralized and decentralized control mechanisms have their own benefits and drawbacks depending on the scale, administrative control, and performance requirements of different types of networks. We would argue that the policy based approach, where different policies dictate different actions to be taken by the network in response to some conditions arising in the network, is an approach that can be used across both paradigms for controlling a network.

With modern AI based capabilities, configuration of different types of software and systems can be done using natural text such as English. Any such tools which provide a natural approach to configure quantum overlays will also fall within this category of tools.

## C. Management Plane Abstraction

Like the control plane architecture, the management plane architecture would need to follow a layered architecture. At each layer in this stack, we assume that the system is conforming to the logical abstractions of a link, a path and a network. This layering of the stack is shown in Fig. 8

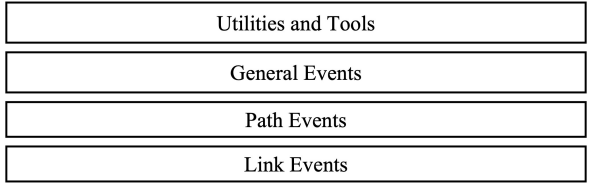

Fig. 8. The Management Plane for a Quantum Network

The different from the control stack and the management stack is the concept of the event. Events happen within the network which need to be handled. These events can then be processed, suppressed, stored, automatically handled by a management software component, or be displayed to a human administrator for action.

The concept of an event is integral to traditional approach for network management, and one can use the existing protocols [40], [41] for network management aspects of a quantum overlay. The format to represent an event, as well as any management information for the link and network are well-defined in those standards. We can extend the specifications to handle the concepts associated with a quantum overlay and a quantum underlay in these specifications.

A big aspect of network management is the use of Artificial Intelligence and Machine Learning to automatically handle the different events that are generated within the network. The concepts of policy based management, described in Section IV-B are useful in network management to handle the events as well. Machine Learning can be used to generate the policies that are required for the handling of the different events in the system [42].

## V. APPLICATIONS

Any discussion of an abstraction is incomplete without a discussion of the applications, i.e. the users of the abstraction that are proposed. In this particular case, we need to discuss the potential applications that can use the abstractions of (i) secure lossy datagram circuit (ii) secure reliable datagram circuit (iii) secure reliable byte-stream and (iv) synchronized random number generator.

### A. Applications using Circuit/ByteStream Abstraction

The first three of these offer abstractions that are similar to those of secure communication pipes that exist in current Internet, with an enhanced security provided by means of the quantum key provided on the quantum network. As a result, the applications of these abstractions will be same as the general class of applications today.

For a secure lossy datagram circuit, real-time applications that can tolerate loss are suitable. These include the streaming of voice and video, conversations over the telephone, and interactive video-conferencing. The primary advantage of the abstraction is that it provides a secure connection by default. In general, any application that is designed around the SIP and RTP protocols on the Internet can benefit from this abstraction, gaining improved security by means of quantum networks.

For a secure reliable datagram circuit, the most common application would be the messaging applications including slack, telegram, whatsapp, facebook messenger, Apple iMessage etc. Since security will come by default for these applications, the secure messaging service which is provided by end-to-end encryption by some of these applications will become available for free. Note however, that these consumer applications are used today on mobile phones which may not have a cost-effective ready access to a free-space quantum communications link, at least at the present. However, the exploitation of this abstraction need not await the development of quantum networking hardware for the hand-held.

There are several enterprise applications which use a messaging protocol between computers in the back-end. The common examples of such messaging protocols include AMQP [43], the original IBM MQSeries [44], various implementations of the concept of the enterprise service bus [45]. These messaging systems operate to interconnect large corporate systems and data centers. An implementation of a quantum network software using the secure reliable datagram circuit would provide commercial use-cases for long-distance quantum networks that are likely to become viable in the near future.

The secure reliable byte-stream is very similar to the well-known Transmission Control Protocol (TCP) which forms the basis for the bulk of communication happening on the Internet. The provision of quantum-security provides a default secure mode of operation for these computers.

### B. Applications using Synchronized Random Number Generator

The synchronized random number generator is a new abstraction that is not found in current software system or communication protocols. We would like to emphasize that the synchronization is reflected in the same random number being visible at all parties that are communicating, not in them happening at the same time concurrently. As a result, this is the abstraction that can lead to the development of new software applications that are enabled purely by the advent of quantum networks. The first and most common use of these synchronized random number is to use as security keys as common shared secrets. However, the ability to generate the same random number at two or more different sites has the potential to improve many other software applications that depend on random numbers.

A good use-case for the random number application can be found in the field of randomized algorithms [46], [47]. There are two broad classes of random algorithms, categorized as Las Vegas algorithms and Monte Carlo algorithms. A Las Vegas algorithm is a randomized algorithm that always gives the correct result but gambles with resources. A Monte Carlo algorithm is a randomized algorithm whose output may be incorrect with some probability. The error probability of a Monte Carlo algorithm would typically be small. Monte Carlo algorithms are used for simulations in a large variety of fields. When using either a Las Vegas or a Monte Carlo algorithm, having two nodes with a synchronized random number available would be an advantage. That allows the algorithm to be parallelized. For Las Vegas algorithms for search, as an example, two or more nodes can use a shared random number to search for items in parallel, resulting in increasing the speed of the search by the number of nodes available.

Monte Carlo simulations [48] can benefit tremendously from a synchronized random number. Monte Carlo simulations of any system depends on drawing random numbers that drives the operations of simulated behavior in a variety of application domains. When the same random number is available at two different sites, Monte Carlo simulations can coordinate their behavior, e.g. use the random number to split the simulation space into two distinct spaces, and each of two computers exploring those two different areas for exploration.

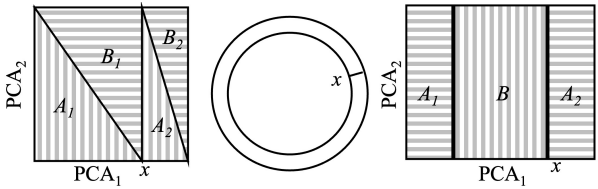

Fig. 9. Different Approaches to Split a Simulation Space

An approach for doing such a split is described below. Let us assume that a Monte Carlo simulation consists of multiple variables. The values these variables can take defines the simulation space that needs to be explored. For the sake of visualization, we can camp these variables into a space of two dimensions, e.g. by using a Principal Component Analysis transform [49]. Without loss of generality, we assume that the axes are bounded by 0 and 1. The random number shared at each of the nodes can be mapped to a number between 0 and 1, e.g. a K bit random number can be taken as a fraction represented by that number divided by $2^K$. This random number can be used to split the space in a manner so that the PCA space is divided into two equal halves. One of the computer can explore Monte Carlo simulation in the simulation space that corresponds to one half of the space split by the random point, while the other can explore the simulation space corresponding to the other half. If the simulation space is mapped to a PCA space of a single dimension, it can be represented as a circular space that loops around after the maximum value, and the random number used to split the space into two equal halves. Many other approaches to split the space in half are also possible, including using the wraparound from the maximum value to separate the 2-dimensional space into two equal halves. These approaches are illustrated in Fig. 9. The left and right pictures in Fig. 9 show two different ways in which a 2-dimensional PCA space can be divided into equal spaces. The middle picture shows how a single dimensional space can be divided into two equal halves by means of a random number. The right picture can be viewed as the unfolded approach to extend the approach shown in the middle from one dimension to two dimensions. Of course, these are only a few approaches to divde the available simulation space, and many other approaches can be designed depending on the exact nature of the variables defining the Monte Carlo simulation.

It is worth noting that the scheme can be easily generalized to split the PCA space and corresponding simulation space into any number of equal partitions, and the simulations run in parallel over several machines. At the end of the simulations, results from all the runs can be aggregated. This random partitioning provides an easier way to coordinate the parallel execution of tasks.

The same splitting of tasks can also be achieved when running other probabilistic algorithms, e.g. training a neural network model in a distributed manner [50]. Neural networks models train themselves by examining different small batches of training data, starting with a random guess of their internal wieghts, and iteratively reducing the error in their prediction by adjusting the weights. If this task has to be conducted in parallel by two or more different machines, they need to cover different batches of the training data in a synchronized manner so they do not use the same data batch, but operate over different data batches. A synchronized random number can provide the seed for them to effectively divide the training space, and reduce the time involved in training the models. This can speed up the exploration of parallel methods for running machine learning models.

Another application area that may benefit is one where data space is not shared among learning models, e.g. when different data fragments are running over multiple wide area nodes.

The current approaches to train models across such distributed data sets without moving data to a central location is federated learning [51], [52] and requires a central coordinator in the most common implementations. Federated learning can be improved significantly by having an exchange of some portion of data among all the participating sites [53]. A random number exchanged among participants can be used to determine what data set to exchange. One nice property about using Quantum networks for this over traditional network is that this synchronization and coordination can be done without the need of a centralized federated learning coordinator.

Other type of applications that can be implemented and accelerated using parallel algorithms include simulated annealing [54] and other genetic algorithms [55]. Since these simulations are done on large compute clusters where the addition of quantum networks would be economically feasible, they provide a good venue to implement applications of quantum communications using the abstractions described in this paper.

As a final note, quantum communications will result in the synchronized number appearing at two or more nodes almost simultaneously with a difference determined by the error in time-synchronization between them. On the other hand, creating a pseudo-random number and sharing it over a classical network incurs a delay of the round-trip latency between the two nodes. Such an exchange on a classical network also becomes complex and cumbersome as the number of nodes involved in creating a shared random number increases. In the field of improving random algorithms, quantum networks may hold a performance advantage over doing similar exchanges on a classical network.

## VI. CONCLUSION

In this paper, we have done an initial exploration of the type of communication abstraction that a quantum communications network can provide to its users, and the potential set of applications that can be implemented using those abstractions. This leads towards the development of a suite of software applications which can lead to practical exploitation of quantum networks. These applications can be developed today with the experimental quantum networks that are under development and deployment.

As a next step in this direction of exploiting quantum networks, we want to explore the abstractions that would make sense when quantum computers are communicating over a quantum network, and when a mixed configuration of classical and quantum computers are used. We would also like to expand the abstractions to a multicast domain, and explore applications when many participants are able to get a synchronized random number at the same time.